\documentclass[reprint,aps,prl,showpacs,eqsecnum,twocolumn,superscriptaddress]{revtex4-1}
\usepackage{amsmath,amssymb,graphicx,color,ulem,multirow,inputenc}

\usepackage[colorlinks=true, linkcolor=blue, citecolor=blue, urlcolor=blue]{hyperref}

\begin{document}


\title{Actinide-boosting $r$ Process in Black Hole--Neutron Star Merger Ejecta}

\author{Shinya Wanajo}
\email{shinya.wanajo@aei.mpg.de}
\affiliation{Max-Planck-Institut f\"ur Gravitationsphysik (Albert-Einstein-Institut), Am M\"uhlenberg 1, D-14476 Potsdam-Golm, Germany}

\author{Sho Fujibayashi}%
\affiliation{Frontier Research Institute for Interdisciplinary Sciences, Tohoku University, Sendai, 980-8578, Japan}
\affiliation{Astronomical Institute, Graduate School of Science, Tohoku University, Sendai, 980-8578, Japan}
\affiliation{Max-Planck-Institut f\"ur Gravitationsphysik (Albert-Einstein-Institut), Am M\"uhlenberg 1, D-14476 Potsdam-Golm, Germany}

\author{Kota Hayashi}
\affiliation{Max-Planck-Institut f\"ur Gravitationsphysik (Albert-Einstein-Institut), Am M\"uhlenberg 1, D-14476 Potsdam-Golm, Germany}
\affiliation{Center for Gravitational Physics and Quantum Information, Yukawa Institute for Theoretical Physics, Kyoto University, Kyoto, 606-8502, Japan}

\author{Kenta Kiuchi}
\affiliation{Max-Planck-Institut f\"ur Gravitationsphysik (Albert-Einstein-Institut), Am M\"uhlenberg 1, D-14476 Potsdam-Golm, Germany}
\affiliation{Center for Gravitational Physics and Quantum Information, Yukawa Institute for Theoretical Physics, Kyoto University, Kyoto, 606-8502, Japan}

\author{Yuichiro Sekiguchi}
\affiliation{Center for Gravitational Physics and Quantum Information, Yukawa Institute for Theoretical Physics, Kyoto University, Kyoto, 606-8502, Japan}
\affiliation{Department of Physics, Toho University, Funabashi, Chiba 274-8510, Japan}

\author{Masaru Shibata}
\affiliation{Max-Planck-Institut f\"ur Gravitationsphysik (Albert-Einstein-Institut), Am M\"uhlenberg 1, D-14476 Potsdam-Golm, Germany}
\affiliation{Center for Gravitational Physics and Quantum Information, Yukawa Institute for Theoretical Physics, Kyoto University, Kyoto, 606-8502, Japan}




\date{\today}

\begin{abstract}
We examine nucleosynthesis in the ejecta of black hole--neutron star mergers based on the results of long-term neutrino-radiation-magnetohydrodynamics simulations for the first time. We find that the combination of dynamical and post-merger ejecta reproduces a solar-like $r$-process pattern. Moreover, the enhancement level of actinides is highly sensitive to the distribution of both electron fraction and the velocity of the dynamical ejecta. Our result implies that the mean electron fraction of dynamical ejecta should be $\gtrsim 0.05$ in order to reconcile the nucleosynthetic abundances with those in $r$-process-enhanced, actinide-boost stars. Since the tidal ejecta preserve the neutron-richness in the inner crust of pre-merging neutron stars, this result provides an important constraint for nuclear equations of state, if black hole--neutron star mergers are responsible for actinide-boost stars.
\end{abstract}

\maketitle


\textit{Introduction.}---One of the long-standing issues in nuclear astrophysics is to determine the mechanism that ensures a robust solar-like $r$-process pattern found in $r$-process-enhanced metal-poor stars (e.g., Ref.~\cite{Cowan2021}). Recent work~\cite{Fujibayashi2023} based on long-term neutrino-radiation-hydrodynamics simulations of neutron star merger remnants suggests that a combination of dynamical and post-merger ejecta leads to a solar-like $r$-process pattern (see also~\cite{Just2015,Martin2015,Kullmann2023}, including cases of black hole--neutron star mergers). 
However, these nucleosynthesis studies are not fully self-consistent such that the three-dimensional simulations for the dynamical phase have been followed by axisymmetric simulations with parameterized viscosity effects for the post-merger phase. 


Another salient feature observed in $r$-process-enhanced stars is the presence of stars with an excess of the radioactive species Th with respect to a stable element, e.g., Eu. This excess is known as the ``actinide boost''~\cite{Cayrel2001,Hill2002,Schatz2002,Wanajo2002}. The Th/Eu ratio ranges over a factor of 8, 
$0.14 \le \mathrm{Th/Eu} \le 1.1$~\cite{Ji2018,Placco2023}. Among those, about one-third of $r$-process-enhanced stars are classified as actinide-boost stars with $\mathrm{Th/Eu} > 0.5$ (the production ratio of $\mathrm{Th/Eu} > 0.9$ assuming a stellar age of 13~Gyr). The study of nucleosynthesis~\cite{Holmbeck2019} using a single thermodynamic trajectory 
has demonstrated that actinides are overproduced if the initial neutron-richness is sufficient for fission recycling. However, they assumed an  equal split of fissioning nuclei, which could result in a significant underestimation of Eu production and thus an overestimation of Th/Eu (see the Supplemental Material \cite{Suppl}).


In this Letter, we present the first nucleosynthesis study based on self-consistent, long-term neutrino-radiation-magnetohydrodynamics simulations in Refs.~\cite{Hayashi2022,Hayashi2022b}. The aim of this work is two-fold. One is to examine if black hole--neutron star mergers can reproduce solar-like $r$-process patterns as can be seen in $r$-process-enhanced stars. 
This is important because of the recent observation of gravitational waves from two compact binary coalescences exhibiting properties consistent with black hole--neutron star mergers \cite{Abbott2021}.
Second we attempt 
to explore the conditions in which actinide boost is realized.
Based on our results, we also provide a constraint for nuclear equations of state.

\textit{Models.}---We adopt the results of neutrino-radiation-magnetohydrodynamics simulations of black hole--neutron star mergers in Refs.~\cite{Hayashi2022,Hayashi2022b}. In their models, the temperature-dependent, tabulated nuclear equations of state DD2~\cite{Banik2014} or SFHo~\cite{Steiner2013} were adopted, and early dynamical and late post-merger mass ejections were self-consistently computed in a single three-dimensional computational domain.
The post-merger mass ejection is predominantly due to the effective viscosity induced by magnetohydrodynamic turbulence.


In this study, we take their models of Q4B5H with DD2 and SFHo (hereafter referred to as Q4B5H-DD2 and Q4B5H-SFHo) and Q6B5L with DD2 (Q6B5L-DD2). 
Here, Q4 and Q6, B5, and H and L stand for the ratios of black-hole masses ($5.4\, M_\odot$ and $8.1\, M_\odot$) with respect to that of a neutron star ($1.35\, M_\odot$), the initial maximum magnetic-field strength ($5\times 10^{16}$~G), and the grid spacing for the finest refinement level (270~m and 400~m), respectively. The ejecta masses were approximately saturated for all models at the end of simulations ($\agt1$--2~s). The total ejecta masses are $M_\mathrm{ej}/M_\odot = 0.070$, 0.031, and 0.067 for Q4B5H-DD2, Q4B5H-SFHo, and Q6B5L-DD2, respectively.


\begin{figure}
\includegraphics[width=0.49\textwidth]{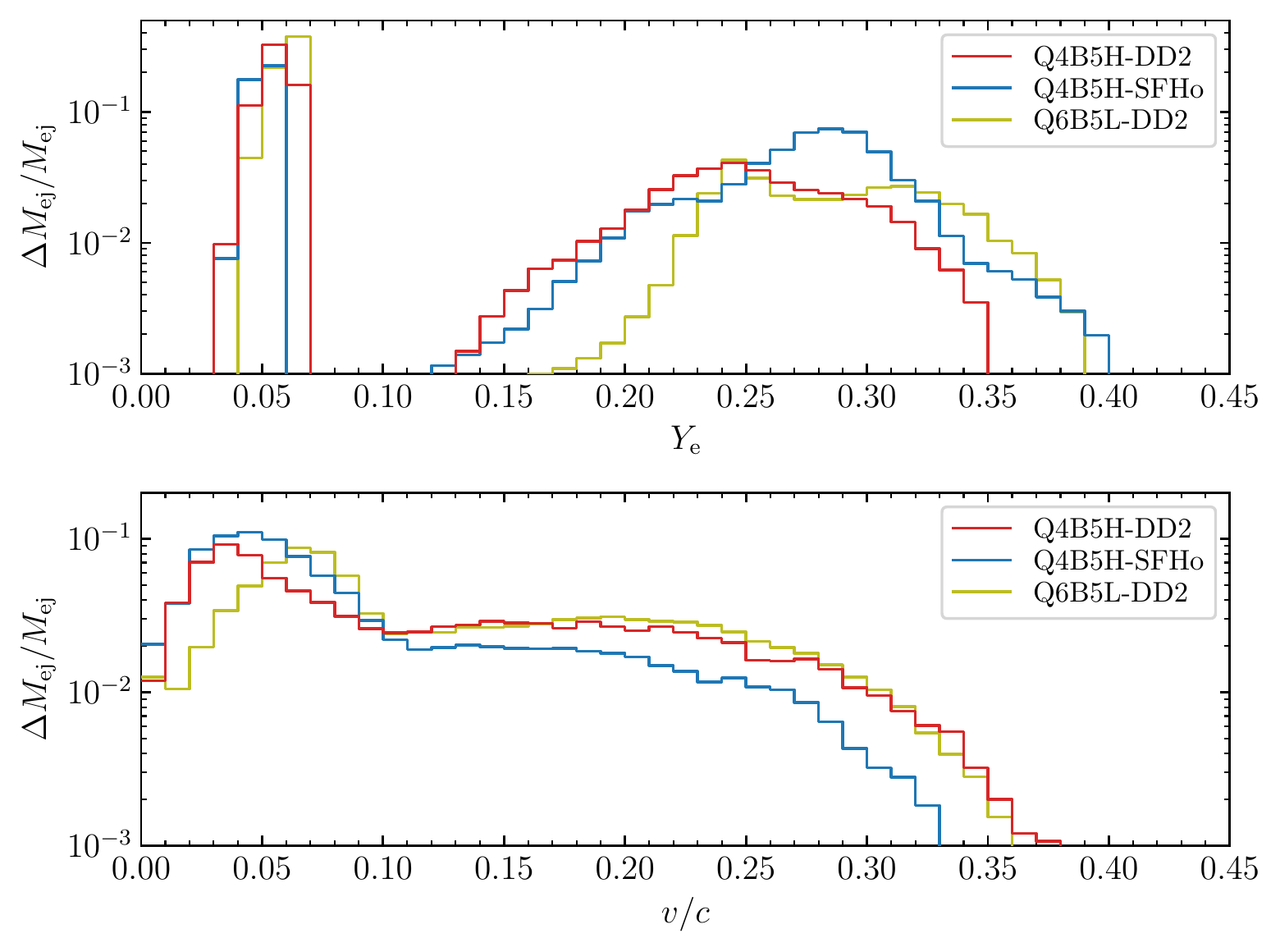}
\caption{\label{fig:histogram}Mass histograms at the end of simulations with respect to the total ejecta masses for $Y_\mathrm{e}$ (top) with an interval of $\Delta Y_\mathrm{e} = 0.01$, and $v/c$ (bottom) with an interval of $\Delta (v/c) = 0.01$. 
}
\end{figure}

For nucleosynthesis calculations during post-processing, about 1400 (of which 400 are for the dynamical component) tracer particles are generated for each model as in Ref.~\cite{Fujibayashi2023}. 
The method of nucleosynthesis calculations is described in the Supplemental Material \cite{Suppl} (see also Refs.~\cite{Wanajo2018,Goriely2008,Tachibana1990,Goriely2010,Goriely2007,Schmidt2010,Timmes2000,Hotokezaka2016,Baym1971} therein). Each nucleosynthesis calculation ends at 1~yr.
Fig.~\ref{fig:histogram} displays the mass histograms for electron fraction (proton number per nucleon) $Y_\mathrm{e}$ and velocity $v$ with respect to the speed of light $c$ (bottom). 
We find both narrow 
and broad distributions for each model at $Y_\mathrm{e} \sim 0.05$--0.06 and 0.1--0.4 originating from dynamical and post-merger ejecta, respectively. The masses of the dynamical ejecta (defined as those with $Y_\mathrm{e} < 0.08$) are $M_\mathrm{ej}/M_\odot = 0.039$ (56\% of the total ejecta mass), 0.012 (39\%), and 0.046 (69\%) for Q4B5H-DD2, Q4B5H-SFHo, and Q6B5L-DD2, respectively. 
We also find narrow and broad distributions at $v/c \sim 0.03$--0.07 and 0.1--0.4, which come from post-merger and dynamical ejecta, respectively.

\textit{$r$-Process Abundance Patterns.}---The nucleosynthesis result for model Q4B5H-DD2 
is presented in Fig.~\ref{fig:abun_comp}.
We find that the dynamical and post-merger ejecta are responsible for the heavy ($A > 130$) and light ($A < 130$) components of $r$-process abundances, respectively. The ensemble of both components is in good agreement with the pattern of $r$-process residuals to the solar abundances ($r$-residuals hereafter)~\cite{Prantzos2020} for $A \approx 90$--210  {(for the comparison with previous studies~\cite{Just2015,Kullmann2023}, see the Supplemental Material \cite{Suppl}). 


\begin{figure}
\includegraphics[width=0.49\textwidth]{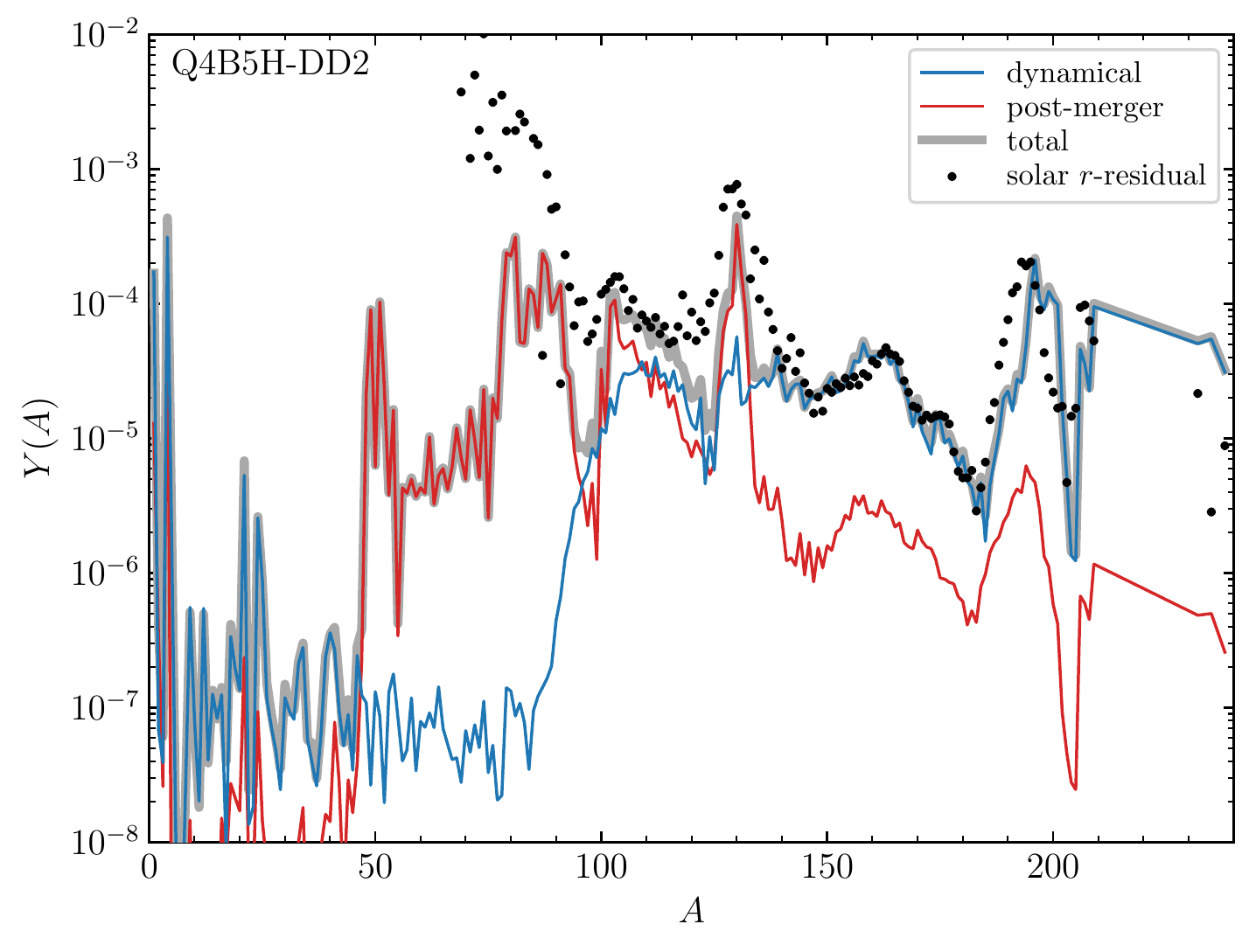}
\caption{\label{fig:abun_comp}Total isobaric abundances (gray) as well as those from dynamical (blue) and post-merger (red) ejecta components at the end of simulation (1~yr; all trans-Pb nuclei except for Th and U are assumed to have decayed) for Q4B5H-DD2. The black circles denote the solar $r$-residuals vertically shifted to match the calculated total abundance of $^{153}$Eu. 
}
\end{figure}

\begin{figure}
\includegraphics[width=0.49\textwidth]{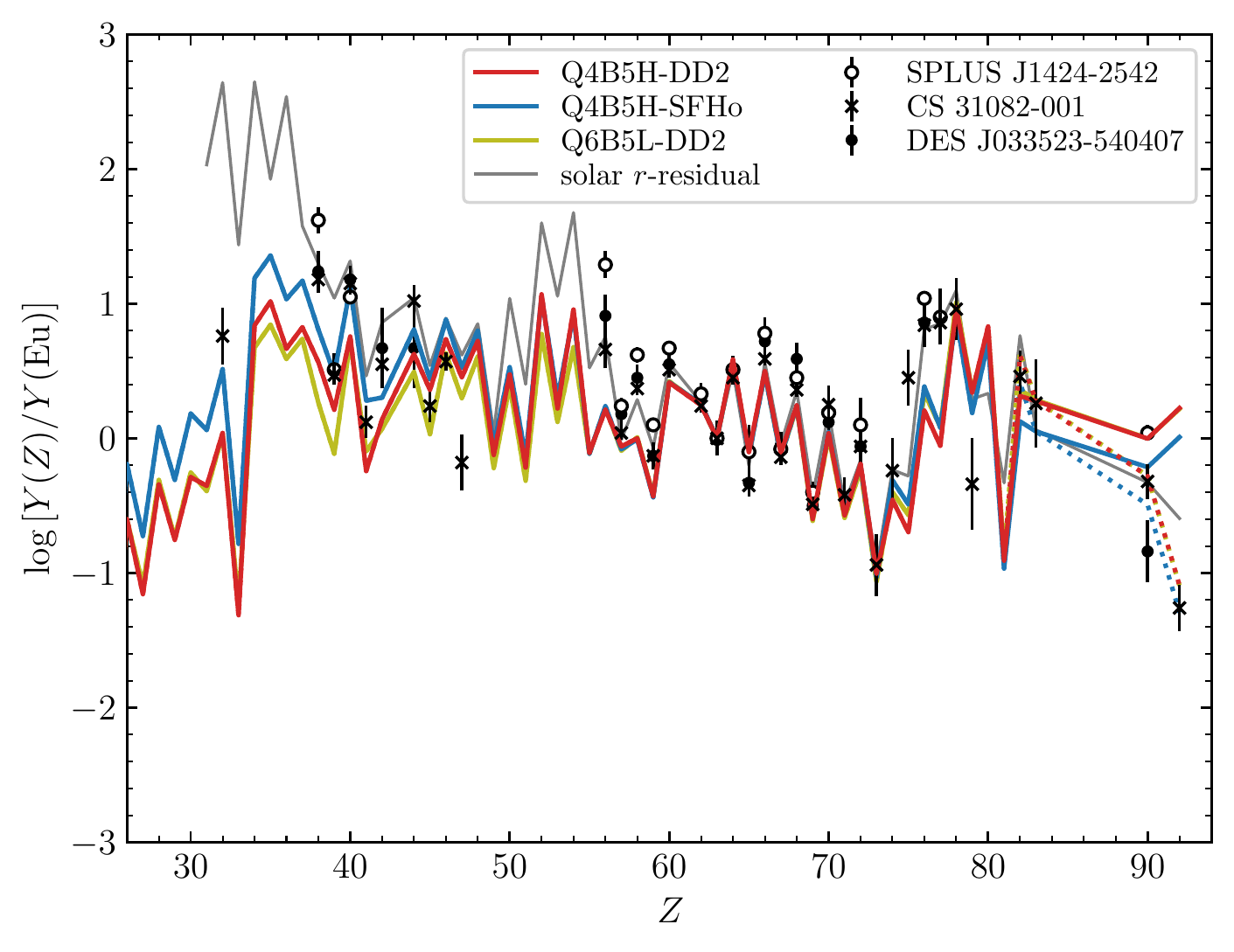}
\caption{\label{fig:abun_model}Elemental abundances normalized by the abundances of Eu for all models. The normalized solar $r$-residuals are also displayed by a gray curve. The dotted lines show the abundances of Pb, Bi, Th, and U at 13~Gyr. The normalized stellar abundances (with error bars) of SPLUS J1424-2542 (open circles), CS~31082-001 (crosses), and DES~J033523-540407 (filled circles) are also shown.}
\end{figure}

\begin{figure*}
\includegraphics[width=0.49\textwidth]{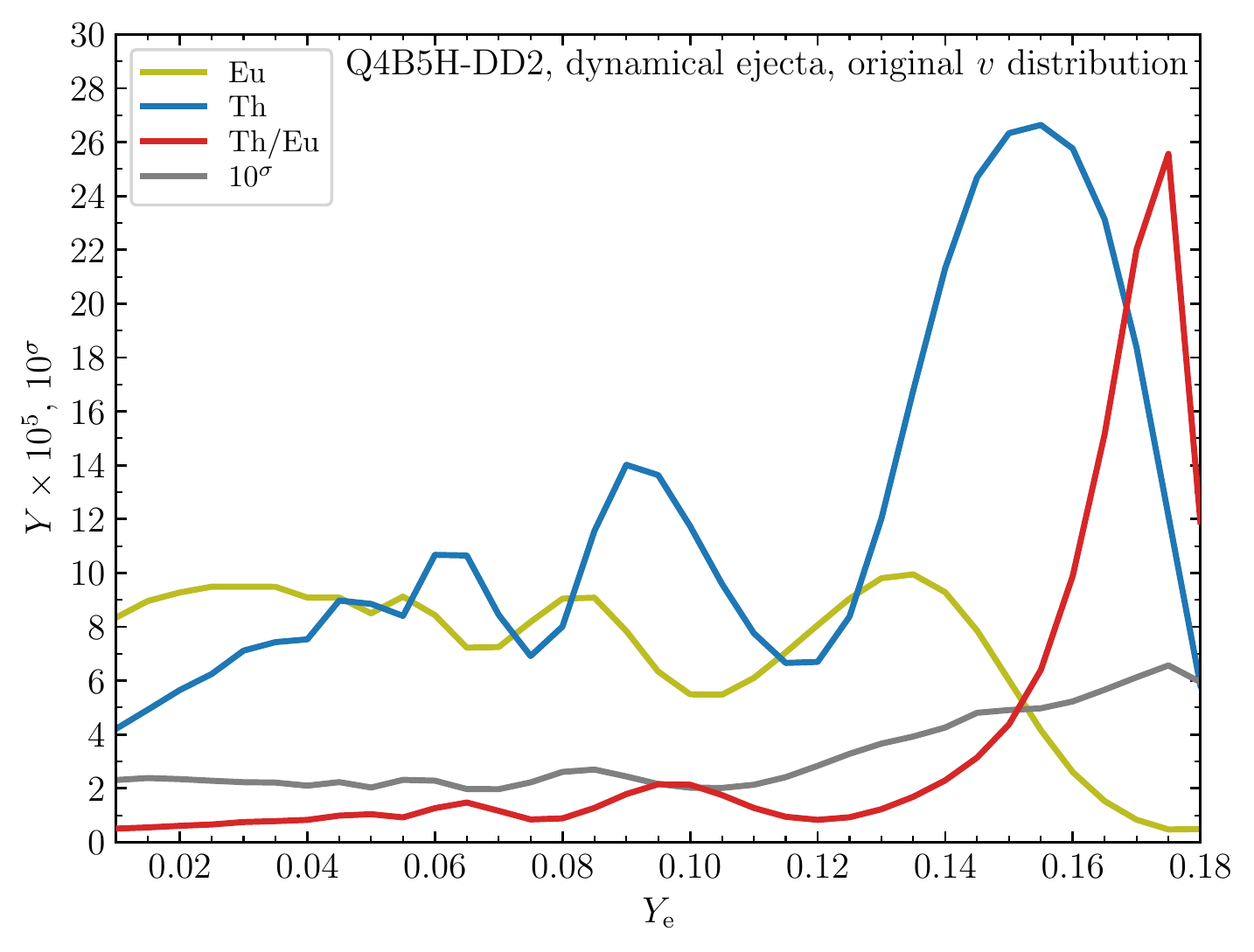}
\includegraphics[width=0.49\textwidth]{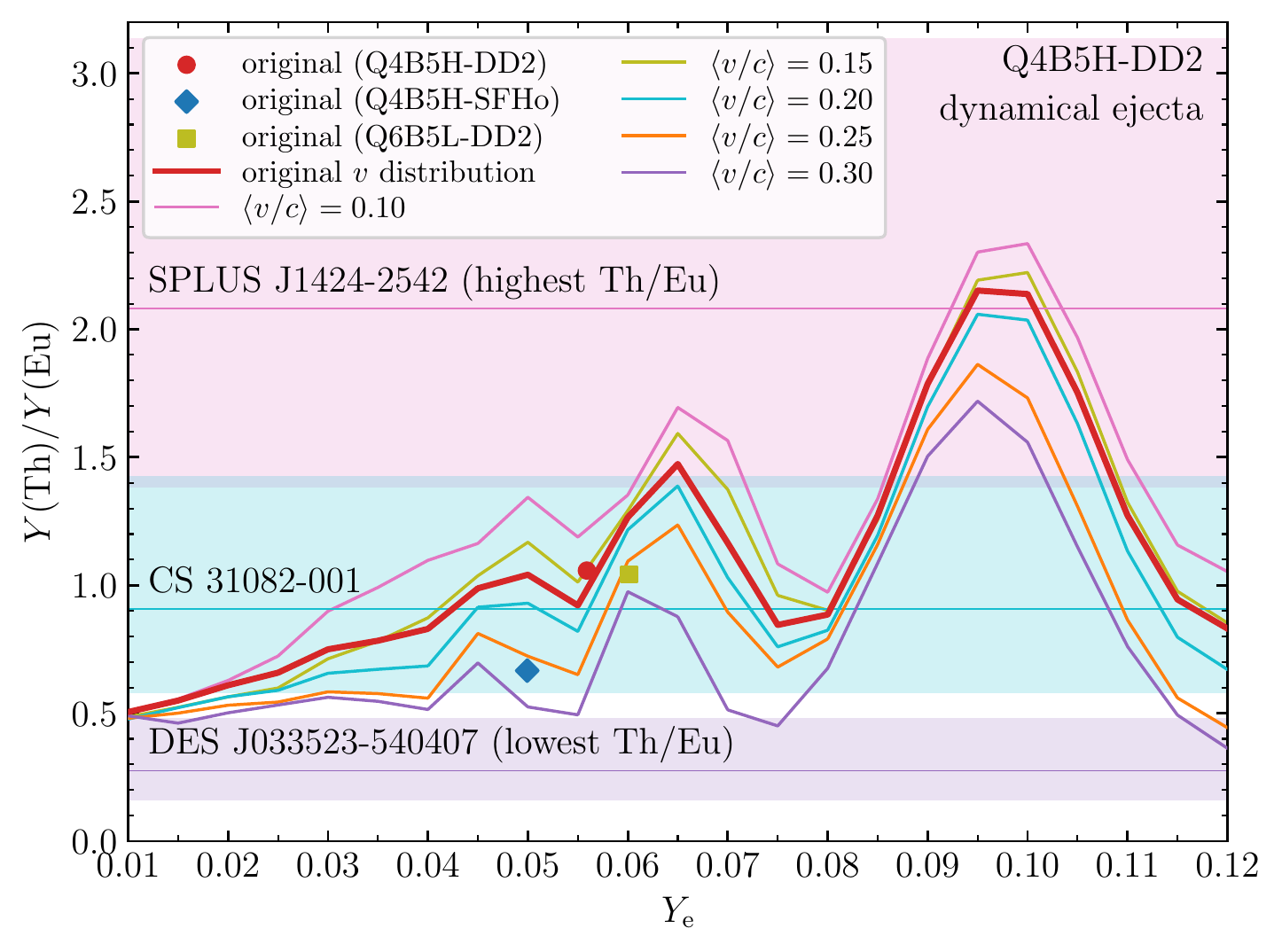}
\caption{\label{fig:theu}Abundances of Eu and Th (at 1~yr; all trans-Pb nuclei except for Th and U are assumed to have decayed) as well as the number ratio Th/Eu (in linear scale) for Q4B5H-DD2 (dynamical ejecta) as a function of $Y_\mathrm{e}$ (see the text). 
In the left panel, the factor $10^\sigma$ is also shown, which indicates the deviation from the solar $r$-residual distribution. In the right panel, Th/Eu ratios calculated using tracer particles with given outflow velocities, mass-averaged in the range of $\pm 0.025$,  are also shown (labelled as $\langle v/c \rangle$). 
The symbols indicate the original Th/Eu ratios for Q4B5H-DD2 (circle), Q4B5H-SFHo (diamond), and Q6B5L-DD2 (square) at each mass-averaged $Y_\mathrm{e}$. The horizontal lines (with errors shown by shaded areas) indicate the observational ratios of SPLUS J1424-2542 (magenta), CS~31082-001 (cyan), and DES~J033523-540407 (purple), where the abundances of Th are corrected to those of 13~Gyr ago.}
\end{figure*}


In the dynamical ejecta, the abundance pattern for $A < 170$ is determined predominantly by the asymmetric fission from nuclei with $A \sim 260$--280 after the end of an $r$~process. This leads to the diminished second peak ($A \sim 130$) as well as the formation of the silver ($A \sim 100$--110)~\cite{Vassh2020} and rare-earth ($A \sim 160$) \cite{Goriely2013} peaks.
It is important to note that the fission properties in the neutron-rich region, including fission fragment distributions, are currently very uncertain \cite{Goriely2015,Vassh2019,Lemaitre2021}. However, we regard the adopted GEF fragment distrubutions \cite{Schmidt2010} as a reasonable choice for the purposes in this study, in which robustly reproducing a solar-like $r$-process pattern for lanthanides, including Eu, is particularly important (for the comparison with the different predictions of fission fragment distributions, see the Supplement Material \cite{Suppl} and Ref.~\cite{Kodama1975} therein). 

Fig.~\ref{fig:abun_model} compares the elemental abundances normalized by those of Eu for all models. It is noteworthy that the abundance patterns of all models are similar for $40 < Z < 82$ ($90 < A < 206$), which are in good agreement with that of the solar $r$-residuals (although the second peak elements are under-produced). 
Changing the binary mass ratio (Q6B5L-DD2) has little impact on the abundance pattern, while adopting the other EOS (Q4B5H-SFHo) leads to about a factor of two higher and lower abundances for the lightest ($Z \le 40$) and heaviest ($Z \ge 82$) elements, respectively.}

Normalized abundances of $r$-process-enhanced stars SPLUS J1424-2542~\cite{Placco2023}, CS~31082-001~\cite{Siqueira2013}, and DES~J033523-540407~\cite{Ji2018} are also plotted. These stars exhibit the highest, high (at the criterion for actinide-boost stars in this study), and lowest values of measured Th/Eu ratio, respectively. 
Note that Th and U are $\alpha$-decaying species (half-lives of 14.05~Gyr, 0.704~Gyr, and 4.47~Gyr for $^{232}$Th, $^{235}$U, and $^{238}$U, respectively), and Pb and Bi are predominantly $\alpha$-decayed products from long-lived progenitors. Thus, the abundances of Pb, Bi, Th, and U at 13~Gyr are also shown by dotted lines, given that these stars were born several 100~Myr after the big bang (e.g., Refs.~\cite{Wanajo2021,Hirai2022,Simon2023}). We find that the abundance patterns are in reasonable agreement with those of the stars. 
In particular, the normalized abundances of Pb, Bi, Th, and U in model Q4B5H-DD2 are in good agreement with those in CS~31082-001, one of the actinide-boost stars.

\begin{figure*}
\includegraphics[width=0.49\textwidth]{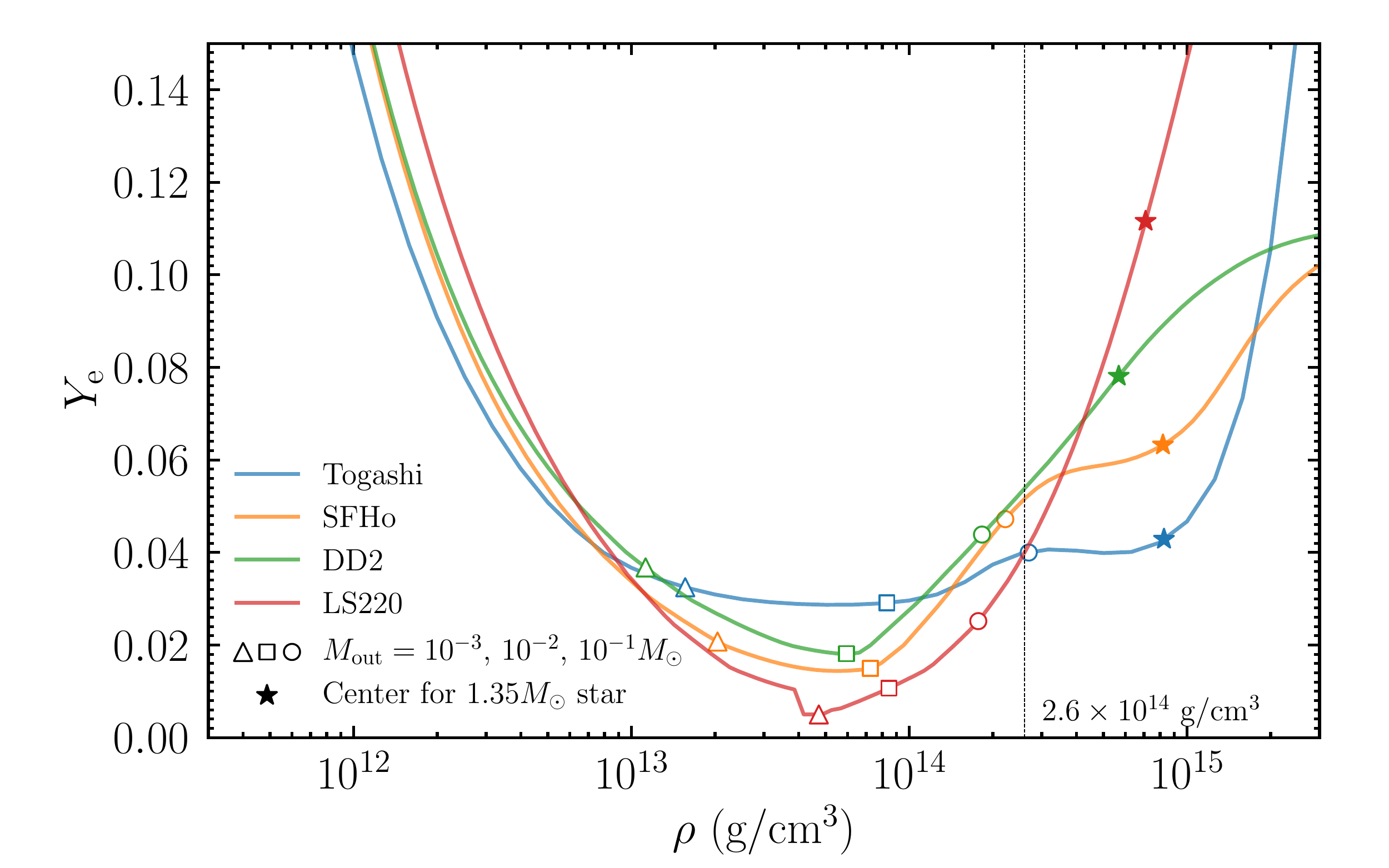}
\includegraphics[width=0.49\textwidth]{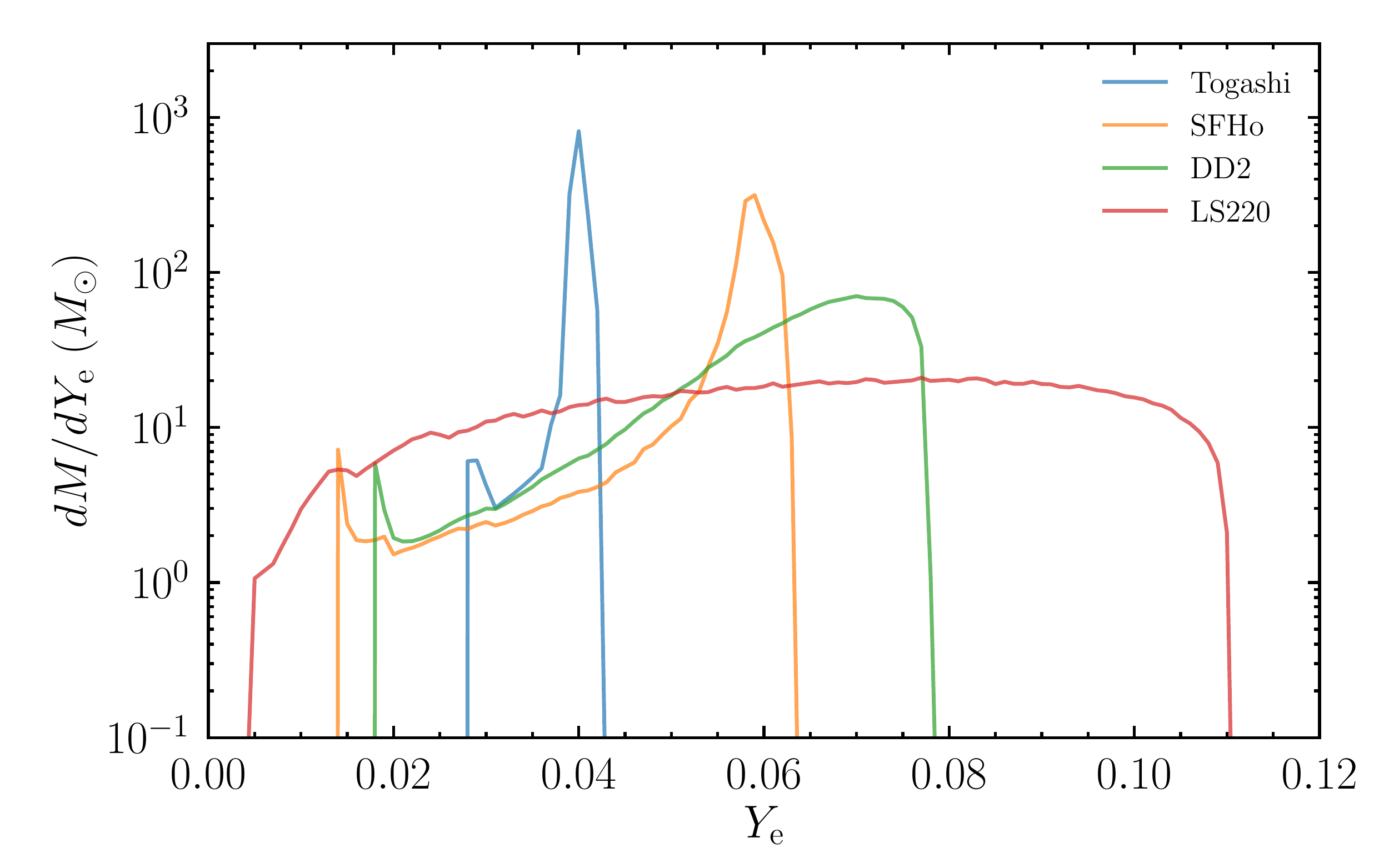}
\caption{\label{fig:eos}Properties of selected EOSs (Togashi, SFHo, DD2, and LS220) under the condition of neutrino-less $\beta$~equilibrium. In the left panel, each curve shows $Y_\mathrm{e}$ as a function of matter density $\rho$ with the central value (star) for the $1.35\, M_\odot$ neutron star. The vertical line indicates the nuclear saturation density. The open triangle, square, and circle symbols mark the masses measured from the surface of the $1.35\, M_\odot$ neutron star, $M_\mathrm{out}/M_\odot = 10^{-3},\, 10^{-2}$, and $10^{-1}$, respectively. The right panel shows the mass distribution of the $1.35\, M_\odot$ neutron star for each EOS as a function of $Y_\mathrm{e}$}.
\end{figure*}

\textit{Actinide Boost.}---Here, we examine the dependencies of actinide production, or Th/Eu, in dynamical ejecta on $Y_\mathrm{e}$ and outflow velocity $v$. The contribution of post-merger ejecta to lanthanide and actinide production is $\sim 0.1$--10\% (Fig.~\ref{fig:abun_comp}), which does not affect our discussion here. The left panel of Fig.~\ref{fig:theu} shows the abundances of Eu and Th (at 1~yr; all trans-Pb nuclei except for Th and U are assumed to have decayed) as well as the ratio Th/Eu for model Q4B5H-DD2 as functions of $Y_\mathrm{e}$. Here, $Y_\mathrm{e}$ is taken to be a free parameter (taking into account possible variation due to the nuclear EOS adopted; see Fig.~\ref{fig:eos}), replacing  the original values by a single $Y_\mathrm{e}$. The Eu abundance becomes minimal at $Y_\mathrm{e} = 0.175$ as the nuclear flow proceeds toward heavier nuclei. With a reduction of $Y_\mathrm{e}$, Eu becomes abundant owing to fission recycling.
The Th abundance reaches a maximum at $Y_\mathrm{e} = 0.155$. For $Y_\mathrm{e} < 0.155$, we find the effects of ``fission waves'' for Eu and Th in response to the second and third fission recycling. 
The multiple fission recycling leads to a convergence of the Th/Eu ratio towards lower $Y_\mathrm{e}$ with a few local maxima (for the comparison with the previous study~\cite{Holmbeck2019}, see the Supplemental Material \cite{Suppl}).


Assuming
that black hole--neutron star mergers are, in part, the sources of $r$-process elements, 
we limit the range of $Y_\mathrm{e}$ for subsequent discussion. The agreement level of the computed abundances with the solar $r$-residuals can be evaluated by the deviation factor $10^\sigma$ with $\sigma$ defined by $\sigma^2 = \sum_A \left[\log_{10} N(A) - \log_{10} N_\odot(A)\right]^2/n_\mathrm{tot}$~\cite{Fujibayashi2023},
where $N(A)$ and $N_\odot(A)$ are the normalized isobaric abundances for the dynamical ejecta and the solar $r$-residuals, respectively. The sum runs over the $n_\mathrm{tot} = 67$ isobars from $A = 139$ to 205, which are predominantly produced in the dynamical ejecta. 
For Q4B5H-DD2, Q4B5H-SFHo, and Q6B5L-DD2 with original $Y_\mathrm{e}$ values, $10^\sigma = 2.2$, 2.1, and 2.0, respectively, indicating agreements with the solar $r$-residual pattern within about a factor of 2. 
Here, we regard the abundance patterns with $10^\sigma < 3$ as appropriate models for $r$-process sources (see also Refs.~\cite{Goriely1999,Fujibayashi2023}). Thus, hereafter we restrict the range of our attention to $Y_\mathrm{e} < 0.12$
.

As can be seen in the right panel of Fig.~\ref{fig:theu}, the Th/Eu ratio (the red-thick curve; Q4B5H-DD2) reaches the local maxima at $Y_\mathrm{e} = 0.050$, 0.065, and 0.095 with an overall decreasing trend twards lower $Y_\mathrm{e}$. 
The plot also displays the Th/Eu ratios for tracer particles with selected outflow velocities. 
The symbols indicate the Th/Eu ratios (at mass-averaged $Y_\mathrm{e}$) for all models, which present the levels of actinide boost with original $Y_\mathrm{e}$ distributions. 

We find substantial variation of the Th/Eu ratio at a given $Y_\mathrm{e}$, with a tendency for this ratio to take on a higher value for a slower ejecta velocity. 
This is due to the fact that the radioactive heating from $r$-processing heats up the initially cold material up to $\sim 0.7$--2\,GK (at which so-called $n\gamma$-$\gamma n$ equilibrium, between neutron capture and its inverse, is established~\cite{Wanajo2007}). Under this condition, the slower (nearly adiabatically expanding) ejecta achieve higher temperature during an $r$~process. At higher temperature, the $r$-process path locates closer to $\beta$~stability (because of $n\gamma$-$\gamma n$ equilibrium), in which $\beta$-decay lifetimes are longer. As a result, on the $r$-process path, more abundances populate in the lighter-mass side ($A < 254$ with few fissile nuclei) than in the heavier-mass side ($A \ge 254$ with predominantly fissile nuclei) at the end of an~$r$ process. The former and the latter include, respectively, the progenitors of Th (by $\alpha$ decay) and Eu (by fission). Thus, the slower ejecta result in a higher Th/Eu ratio, although such a dependency diminishes as $Y_\mathrm{e}$ approaches 0.01 because of the multiple fission recycling (that leads to nearly a static nuclear flow).


Higher-$Y_\mathrm{e}$ ejecta (with a fixed velocity) also achieve higher temperature because of an initially larger amount of heavy nuclei ($^{118}$Kr in this study; see the Supplemental Material \cite{Suppl}) and thus higher radioactive energy from $r$-processing. Thus, higher-$Y_\mathrm{e}$ ejecta tend to result in a higher Th/Eu ratio owing to the $r$-process path closer to $\beta$~stability, although the effect of fission waves appears to dominate for $Y_\mathrm{e} > 0.05$.
It is important to note that the Th/Eu ratio depends in particular on theoretical $\beta$-decay predictions~\cite{Holmbeck2019} (see other nuclear ingredients that affect the actinide production in Refs.~\cite{Holmbeck2019,Kullmann2023}). 
The adopted $\beta$-decay rates (GT2~\cite{Tachibana1990} based on the HFB-21 mass prediction~\cite{Goriely2010}) appear to result in a relatively high Th/Eu ratio according to Ref.~\cite{Kullmann2023}.

The stellar Th/Eu ratios of SPLUS J1424-2542~\cite{Placco2023}, CS~31082-001~\cite{Siqueira2013}, and DES~J033523-540407~\cite{Ji2018} are also presented by horizontal lines in the right panel of Fig.~\ref{fig:theu}, where the observational values of Th are corrected to those 13~Gyr ago (a few~Gyr of difference does not substantially change our conclusion). The errors for Th/Eu shown by shaded areas ($\pm 0.18$~dex, $\pm 0.20$~dex, and $\pm 0.24$~dex, respectively) are the root-mean-squares of those for Th and Eu. We find that the models with DD2 result in actinide boost, i.e., Th/Eu $> 0.9$, which reconcile with the Th/Eu ratio of an actinide-boost star CS~31082-001 but not the highest measured Th/Eu ratio of SPLUS J1424-2542. 
The model with SFHo does not meet the condition for the actinide boost, although the Th/Eu ratio resides within the range of the error for CS~31082-001.

According to the results with single $Y_\mathrm{e}$ values 
for Q4B5H-DD2 (with the original velocity distribution, red curve), only a range of $Y_\mathrm{e} \approx 0.045$--0.070 (or 0.080--0.115, which is unlikely according to available EOSs) meets the level of actinide boost, Th/Eu $> 0.9$, or $Y_\mathrm{e} \sim 0.065$ (or 0.085--0.110) for the most actinide-boosted star SPLUS J1424-2542. 
This result can potentially serve as an important constraint for nuclear EOSs, since the dynamical ejecta of black hole--neutron star mergers are expected to preserve the original $Y_\mathrm{e}$ in the inner crust of neutron stars (for the difference from the case of binary neutron star mergers, see the Supplemental Material \cite{Suppl} and Ref.~\cite{Abbott2021b} therein).

The left panel of Fig.~\ref{fig:eos} displays $Y_\textrm{e}$ as a function of matter density under the condition of neutrino-less $\beta$~equilibrium for each EOS. Here, other two EOSs, LS220~\cite{Lattimer1991} and Togashi~\cite{Togashi2017}, are also included (the tables are provided by CompOSE \cite{Typel2015,Oertel2017,Typel2022}; \url{https://compose.obspm.fr}). For Togashi, SFHo, LS220, and DD2, the radii (km) of the $1.35\, M_\odot$ neutron star and maximum masses ($M_\odot$) are (11.6, 2.21), (11.9, 2.06), (12.7, 2.06), and (12.8, 2.42), respectively. For a given EOS, $Y_\mathrm{e}$ increases at high density ($>$ several $10^{13}$~g/cm$^3$) in the presence of symmetry energy. By comparing the curves for DD2 and SFHo with Fig.~\ref{fig:histogram} (top), we find that the matter slightly above the saturation density is tidally ejected (note that the mass below $\sim 10^{13}$~g/cm$^{-3}$ with $Y_\mathrm{e} > 0.04$ is subdominant compared to the typical tidal ejecta mass as indicated by open triangles). 
Given that the matter with a similar density is ejected for other EOSs, the possible range in the tidal ejecta is expected to be $Y_\mathrm{e} \sim 0.04$--0.06, of which the Togashi EOS does not meet the condition for the actinide boost. As can be seen from the right panel of Fig.~\ref{fig:eos}, the neutron star constructed with the Togashi EOS does not contain the matter with $Y_\mathrm{e} > 0.04$, which cannot explain the actinide boost regardless of hydrodynamical conditions.


\textit{Conclusion.}---We have conducted the first exploration of nucleosynthesis based on self-consistent magnetohydrodynamics simulations of black hole--neutron star mergers~\cite{Hayashi2022,Hayashi2022b}. Lighter ($A < 130$) and heavier ($A > 130$) $r$-process nuclei were synthesized in the early dynamical and late-time post-merger components, respectively, the ensemble of these reproducing solar-like $r$-process patterns. This indicates that in addition to binary neutron star mergers, black hole--neutron star mergers can also be galactic $r$-process sites.

Our result has demonstrated that the presence of actinide-boost stars
can be explained if the range for the bulk $Y_\mathrm{e}$ in the dynamical ejecta of black hole--neutron star mergers is $\gtrsim 0.05$ (provided that the trend of Th/Eu for DD2 as a function of $Y_\mathrm{e}$ is similar for other EOSs). 
This range of $Y_\mathrm{e}$ can be an important constraint on nuclear EOSs, such as symmetry energy, under the assumptions that the black hole--neutron star mergers are responsible for actinide-boost stars and the matter slightly above the saturation density is tidally ejected. 
Our result supports a DD2-like EOS, while those without $Y_\mathrm{e} \gtrsim 0.05$ components in the neutrons stars are disfavored as to be realistic equations of state, although we should keep in mind potential changes due to uncertainties in relevant nuclear ingredients.

\textit{Acknowledgement.}---We thank L. Held for his careful reading of the manuscript. We also thank T. Hatsuda for fruitful discussion on nuclear EOSs. This work was in part supported by Grant-in-Aid for Scientific Research (Grant Nos.~JP20H00158 and 23H04900) of Japanese MEXT/JSPS.  Numerical computations were performed on Cobra, Raven and Sakura clusters at Max Planck Computing and Data Facility and on Yukawa21 at Yukawa Institute for Theoretical Physics, Kyoto University.

\newcommand{\aap}{Astron. Astrophys.}
\newcommand{\apjl}{Astrophys. J. Lett.}
\newcommand{\apjs}{Astrophys. J. Suppl.}
\newcommand{\apss}{Astrophys. Space Sci.}
\newcommand{\mnras}{Mon. Not. Roy. Astron. Soc.}
\newcommand{\nphysa}{Nucl. Phys. A}

\bibliography{reference}


\clearpage

\noindent\textbf{Supplemental Material}

\noindent
We present supplemental information to the main text, namely, the method of nucleosynthesis calculations, the comparison of abundance patterns with different fission fragment distributions,  the comparison of our results with the previous relevant studies, and the difference from the case of binary neutron star mergers.

\textit{Method of Nucleosynthesis Calculations.}--- Nucleosynthetic abundances in each tracer particle are computed in post-processing using the nuclear reaction network code \texttt{rNET} described in Ref.~[19]. 
Specifically, theoretical neutron-capture (TALYS [20]) and $\beta$-decay (GT2 [21]) rates are those obtained based on the HFB-21 mass prediction [22]. Experimentally evaluated rates for fission are taken from Nuclear Wallet Cards (\url{http://www.nndc.bnl.gov/wallet/}). Theoretical fission rates based on the HFB-14 mass model [23] are adopted when the experimental data are not available.
Fission fragment distributions as well as the number of prompt neutrons per event are computed with the code of the GEF model [24] (version 2021/1.1; \url{http://www.khschmidts-nuclear-web.eu/GEF-2021-1-1.html}) for spontaneous and  neutron-induced channels, with the latter assuming an incident neutron energy of 0.1~MeV. For the $\beta$-delayed channel, the fragment distributions
for the neutron-induced fission are adopted.

Thermodynamic histories are determined based on matter density in tracer particles. The time evolution of the temperature is computed with the chemical composition, density, and entropy by using the equations of state for a gas consisting of non-degenerate ions, arbitrarily degenerate, relativistic electrons and positrons, and photons [25]. The energy feedback from $\beta$ decay, $\alpha$ decay, and fission is reflected in the increase of entropy at each time step, where 40\% of the energy per $\beta$ decay is assumed to be lost as free-streaming neutrinos [26].

The initial compositions for dynamical and post-merger ejecta are treated separately as follows. For the dynamical (or cold) ejecta, the initial condition is set at the neutron-drip density $\rho_\mathrm{drip} = 4\times 10^{11}$~g~cm$^{-3}$. The initial entropy is taken at this density in a given tracer particle (typically less than 1~$k_\mathrm{B}$/nucleon, where $k_\mathrm{B}$ is Boltzmann constant). The matter is assumed to consist of free neutrons and single representative heavy nuclei of $(Z, A)$ with the numbers per nucleon of $Y_\mathrm{n}$ and $Y(Z, A)$, respectively. These values satisfy mass conservation, $1 = Y_\mathrm{n} + A Y(Z, A)$,
as well as charge neutrality, $Y_\mathrm{e} = Z Y(Z, A)$.
We further assume that the free energy of cold ($\ll 5$~GK) tidal ejecta is given solely by the mass of species per nucleon, i.e.,
\begin{eqnarray}
    f(Z, A)/c^2 & = & m_\mathrm{n} Y_\mathrm{n} + m(Z, A) Y(Z, A) \nonumber \\
                & = & m_\mathrm{n} (1 - A Y_\mathrm{e}/Z)+ m(Z, A) Y_\mathrm{e}/Z,
                \label{eq:free}\nonumber
\end{eqnarray}
where $m_\mathrm{n}$ and $m(Z, A)$ are the masses of a neutron and a nucleus $(Z, A)$, respectively. For a given $Y_\mathrm{e}$, minimization of $f(Z, A)$ gives the representative $(Z, A)$. We find that the nucleus is uniquely determined to be $(Z, A) = (36, 118)$ at the shell closure of $N = 82$, i.e., $^{118}$Kr, for $Y_\mathrm{e} < 0.3$ with the nuclear masses of HFB-21. Note that the equations of state adopted in this study also give the same representative nucleus (see also the same result in Ref. [27]). 

For post-merger (or hot) ejecta, each calculation is initiated from the density at which the temperature decreases to 10~GK in a given tracer particle. The initial compositions are set to $1 - Y_\mathrm{e}$ and $Y_\mathrm{e}$ for free neutrons and protons, respectively, where $Y_\mathrm{e}$ here is taken from the final value in a given tracer particle. At such high temperature, the compositions immediately relax to those in nuclear statistical equilibrium.

\textit{Comparison with Different Fission Fragment Distributions.}---Fig.~\ref{fig:abun_fission} compares the abundances in dynamical ejecta for Q4B5H-DD2 with different fission-fragment distributions of GEF, KT75 (phenomenological double-Gaussian distributions in Ref.~[34]), and KT75-symmetric (assuming symmetric fission with single-Gaussian distributions). We find that the total abundances for lanthanides with $A < 170$ are entirely determined by the fission-fragment distributions adopted. While the GEF prediction leads to an excellent match of the lanthanide distribution to that of the solar $r$-residuals, no agreement can be seen when the KT75 or KT75-symmetric distributions are adopted. 
Thereore, we adopt the GEF fission fragment distributions in this study.

\renewcommand{\thefigure}{S\arabic{figure}}
\setcounter{figure}{0}
\begin{figure}
\includegraphics[width=0.49\textwidth]{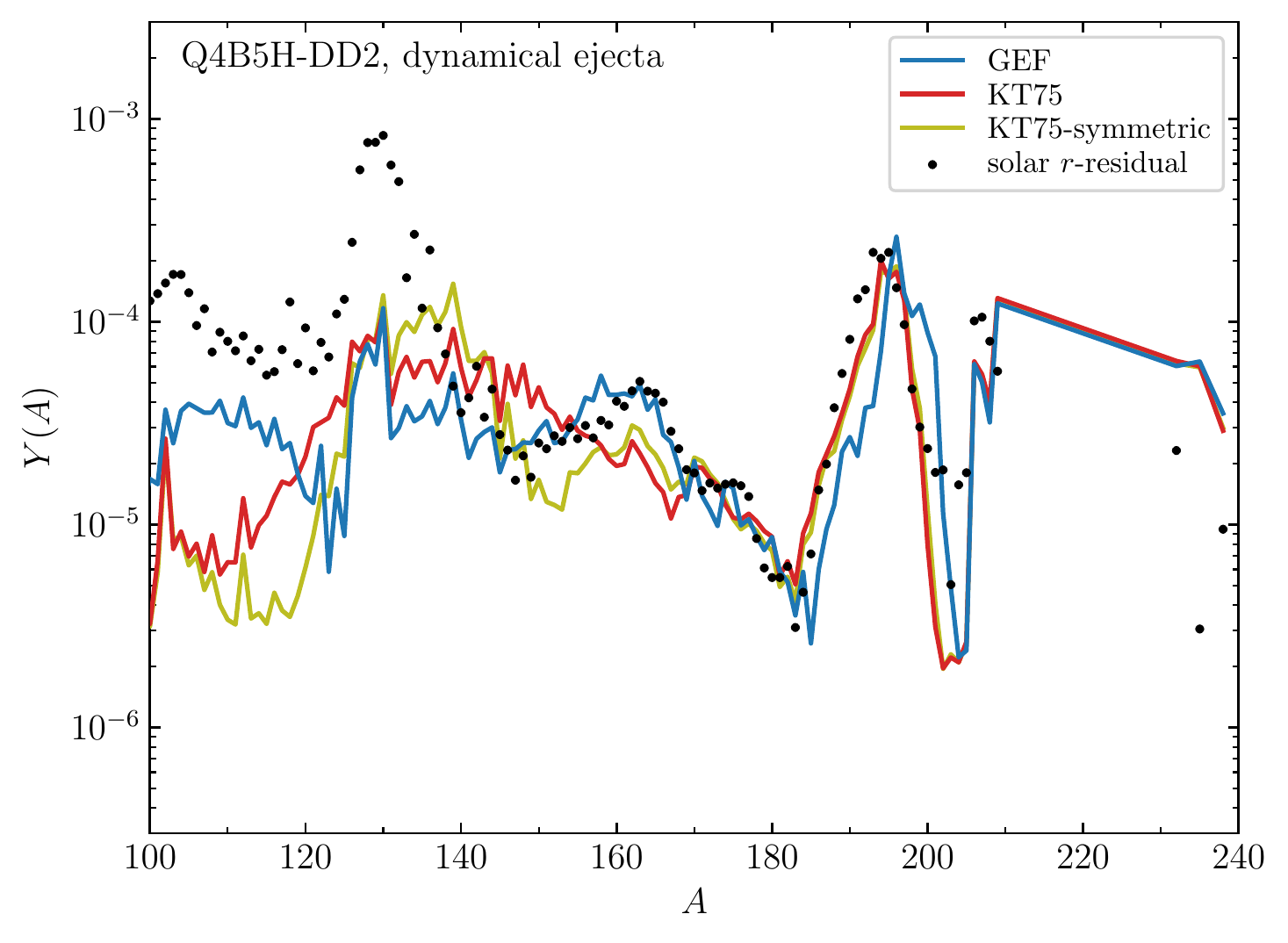}
\caption{\label{fig:abun_fission}Comparison of isobaric abundances from dynamical ejecta with different fission-fragment distributions of GEF (blue), KT75 (red), and KT75-symmetric (olive) at the end of simulation (1~yr; all trans-Pb nuclei except for Th and U are assumed to have decayed) for Q4B5H-DD2. The black circles denote the solar $r$-residuals vertically shifted to match the calculated total abundance of $^{153}$Eu for the GEF case. }
\end{figure}

\textit{Comparison with Previous Studies.}---Our nucleosynthesys result (Fig.~\ref{fig:abun_comp}) is in qualitative agreement with those in Refs.~[3, 5], in which the combined abundances of the dynamical and post-merger ejecta lead to a good agreement with the solar $r$-residual pattern. 
However, in their results [3, 5] the role of the post-merger ejecta to the heavy nuclei ($A > 130$) production substantially varies (from negligible to comparable contributions) depending on the prescription of parameterized viscosity heating. 
Also, these studies adopted simplified initial conditions for the post-merger evolution instead of using the outputs from the early dynamical phase.
In our result, since the effective viscosity heating is due to (parameter-free) magnetohydrodynamic turbulence, the contribution of the post-merger ejecta to the heavy ($A > 130$) component is uniquely determined to be $\sim 0.1$--10\%.

The behaviors of Eu and Th production as a function of $Y_\mathrm{e}$ (Fig.~4) are in qualitative agreement with the result in Ref.~[12] that adopts a single thermodynamic trajectory with conditions relevant to binary neutron stars. However, some quantitative differences from ours can be seen such as the locations of local maxima as well as the values of Th/Eu ratios. In particular, their assumption of the symmetric (equal split) fission results in the population of fragments at $A \sim 130$--140 and thus about a 10 times smaller amount of Eu  $A = 151$ and 153; see their Fig. 3) than that in our case (Fig. 4, left). This leads to too high Th/Eu ($> 3$, which is above that of the most actinide-boosted star; Fig.~4, right) in all the range of $Y_\mathrm{e} < 0.15$ (see also an overall under-production of lanthanides with $A < 170$ for our KT75-symmetric case in Fig.~\ref{fig:abun_fission}). 
Therefore, the high Th/Eu ratios found in Ref. [12] may be due to the underestimation of Eu (i.e., not necessarily due to actinide boost) as a result of adopting the simplistic fission-fragment distribution. In fact, recent studies adopting a more realistic fission model (refs. [5, 29-33]) indicate that the asymmetric fission leads to the fragment population in the lanthanide region including Eu as found in our result. It should be also noted that,
since the abundances of all $r$-process-enhanced stars exhibit remarkable agreements with the solar $r$-residual pattern (i.e., without an under-production of Eu), it is crucial to utilize the fission fragment distribution that leads to a reasonable agreement with the solar pattern including Eu for our purpose.

\textit{Difference from Neutron Star Mergers.}---Unlike binary neutron star mergers [2], thermal effects 
by shock heating are unimportant in the cold tidal ejecta (i.e., without weak interaction) of black hole--neutron star mergers. This fact drastically simplifies the physical conditions relevant to nucleosynthesis, and also enables us to testify the neutron-richness in the inner crust of a neutron star.

The mergers of asymmetric binary neutron stars also tidally eject (marginally weak-processed) neutron-rich material sufficient for fission recycling. In fact, the result in Ref. [2] indicates that more asymmetric-mass binaries lead to higher Th/Eu ratios. However, their asymmetric-mass models (adopting the SFHo EOS) do not result in the actinide boost (up to Th/Eu $\sim 0.45$), probably due to the lower bulk $Y_\mathrm{e} \sim 0.04$ in the tidal ejecta stemming from the less dense region below the saturation density (Fig.~5, left). Therefore, it appears that the mergers of black hole--neutron star binaries provide more favorable conditions for actinide-boosting $r$ process.

One possibility is that black hole--neutron star mergers are responsible for all actinide-boost stars, while binary neutron star mergers account for the rest (about 2/3) of $r$-process-enhanced stars. In fact, the estimated event rate of black hole--neutron star mergers from the gravitational-wave observation is $45^{+75}_{-33}$~Gpc$^{-3}$~yr$^{-1}$ or $130^{+112}_{-69}$~Gpc$^{-3}$~yr$^{-1}$ [16] (depending on the assumption of component-mass distributions) that can be up to about 1/3 of that of binary neutron star mergers, $320^{+490}_{-240}$~Gpc$^{-3}$~yr$^{-1}$ [41], although the uncertainty is quite large.

\end{document}